\documentstyle{mn}
\include{psfig}
\begin{document}
\newcommand{\Ha}{H$\alpha$}
\newcommand{\Ham}{H\alpha}
\newcommand{\Hbm}{H\beta}
\newcommand{\Hb}{H$\beta$}
\newcommand{\oiii}{{\rm \mbox{[O \sc iii]}\ }}
\newcommand{\oii}{{\rm \mbox{[O \sc ii]}\ }}
\newcommand{\nii}{{\rm \mbox{[N \sc ii]}\ }}
\newcommand{\sii}{{\rm \mbox{[S \sc ii]}\ }}
\newcommand{\niioiii}{{\rm \mbox{[N \sc ii]}}/{{\rm \mbox{[O \sc iii]}}\ }}
\newcommand{\oiiinii}{{\rm \mbox{[O \sc iii]}}/{{\rm \mbox{[N \sc ii]}}\ }}
\newcommand{\niioii}{{\rm \mbox{[N \sc ii]}}/{{\rm \mbox {[O \sc ii]}}\ }}
\newcommand{\siioii}{{\rm \mbox{[S \sc ii]}}/{{\rm \mbox {[O \sc ii]}}\ }}
\newcommand{\OiiOiii}{{\rm \mbox{[O \sc ii]}}/{{\rm \mbox {[O \sc iii]}}\ }}
\newcommand{\ohb}{{\rm \mbox{[O \sc iii]/{H$\beta$}}\ }}
\newcommand{\oiihb}{{\rm \mbox{[O \sc ii]/{H$\beta$}}\ }}
\newcommand{\oiioiii}{{\rm ([O \sc ii] + [O \sc iii])/{H$\beta$}\ }}
\newcommand{\hahb}{H$\alpha$/H$\beta$}
\newcommand{\nha}{{\rm \mbox{[N \sc ii]/H$\alpha$}\ }}
\newcommand{\hii}{{H~\sc ii}\ }
\newcommand{\HII}{{H~\sc ii}\ }
\newcommand{\hi}{{H~\sc i}\ }
\newcommand{\om}{{\rm \mbox{[O \sc iii]}}}
\newcommand{\nm}{{\rm \mbox{[N \sc ii]}}}
\newcommand{\kms}{\rm km s$^{-1}$}

\title[O/H in NGC 1365]{The Abundance Gradient of NGC 1365: Evidence
for a Recently Formed Bar in an Archetype Barred Spiral Galaxy?}

\author[Roy \& Walsh]
{J.-R. Roy$^1$ \& J. R. Walsh$^2$\\
$^1$ D\'epartement de physique and Observatoire du mont M\'egantic,
Universit\'e Laval, Qu\'ebec Qc G1K 7P4, Canada\\
$^2$ European Southern Observatory, Karl-Schwarzschild-Strasse 2,
D-85748 Garching bei M\"unchen, Germany}

\maketitle

\begin{abstract}
Emission-line optical spectrophotometry for 55 \hii regions
in the prominent southern barred spiral galaxy NGC 1365 is presented.
Nebular diagnostic diagrams such as \niioii and \siioii versus \oiioiii
show that the \hii regions of the barred galaxy have the same
range of physical conditions as found in non-barred late-type galaxies. Extinction
is moderately high across the disc and there is evidence for a
slight trend of extinction with galactocentric distance; the
logarithmic extinction at \Hb\  falls from about  $c$(\Hb) = 1.2
in the centre to 0.6 -- 0.8 in the outer regions. The global O/H
distribution has a moderate gradient of $\sim -0.5$ dex $\rho_0$$^{-1}$
($\sim -0.02$ dex kpc$^{-1}$) consistent with the known
trend between the slope of the abundance gradient and
the strength of the bar. 
A break is seen in the O/H gradient just beyond the -4/1 resonance, the gradient being
moderately steep at $\sim -0.8$ dex $\rho_0^{-1}$ (--0.05 dex/kpc) 
inside this resonance, and flat
beyond $\rho/\rho_0$ $\geq$ 0.55. The abundance distribution is compared
with another barred spiral galaxy, NGC 3359,
and with that of two well-sampled normal spiral galaxies, NGC 2997
and M~101. The possibility that the bar formed recently in NGC 1365 is
considered. The difficulties encountered
in doing spectrophotometry with fibre optics are discussed and shown not
to be insurmountable.
\end{abstract}

\section{Introduction}

Bars, or non-axisymmetric central light distributions, are found in about
2/3 of the disc galaxies in the present-day universe (Sellwood \&
Wilkinson \shortcite{SW93}).
 Mechanisms for bar formation have been explored quite extensively
(Noguchi \shortcite{NO88},  \shortcite {NO96A}, \shortcite
{NO96B}, Shlosman \& Noguchi \shortcite{SN93}). As a result of cooling, gas discs are subject
to gravitational instability and bars can arise spontaneously.
In galaxies of earlier types, due the stabilizing effect of the
growing bulge, the disc is much less prone to spontaneous
bar formation. A stronger perturbation is then needed, like an interaction
with a companion, a merger or the tidal forces of  a galaxy cluster.
The bar will not last forever because it may dissolve 
when the core mass becomes important, and will contribute to
the growth of the bulge (\cite{NS96}). This implies that some bars are young, and
that some bulges could contain relatively young stars.

The fact that the heavy element abundance distribution
is flatter in barred galaxies as compared with normal spirals of
similar type can be explained by the action of inward
and outward radial flows of interstellar gas induced by the non-axisymmetric potential
of bars (Roberts et al. \shortcite{RO79}; Sellwood \& Wilkinson \shortcite{SW93};
Friedli et al. \shortcite{FR94}). As a consequence,
the slope of any pre-existing radial
gaseous or stellar abundance gradient decreases
with time. The long term effect is such
that the stronger the bar is, the flatter the abundance
gradient becomes with time (Friedli \& Benz \shortcite{FB95}). The finding
by Martin \& Roy \shortcite{MR94} of a relationship between the slopes of the global
 abundance gradient and the strength of bars provides support for the scenario
of recently formed bars (Roy \shortcite{RO96B}). 
Thus bars may not necessarily  be primordial
features (Combes \& Elmegreen \shortcite{CE93}), but could also form at any time
during the lifetime of a galaxy (Friedli \& Benz \shortcite{FB95}; 
Martinet \shortcite{MAR95}; Martin \& Roy
\shortcite{MR95}).

In late-type galaxies (SBc, SBb), a young bar ($\leq$1 Gyr) will be characterized as being
a gas-rich structure. Such bars can be the site of vigorous  
star formation;
if the star formation process is not inhibited by the high cloud velocities,
the radial abundance distribution will present
a steep inner gradient (homogenizing effects being compensated by
chemical enrichment due to vigorous star formation),
combined with a flatter gradient beyond corotation  due
to dilution by the outward radial flows, as shown by Friedli et al. \shortcite{FR94}
and Friedli \& Benz \shortcite{FB95}.
On the other hand, if the level of star formation in the bar is modest or absent
due to inhibiting forces, the radial abundance gradient
would be weak both in the bar and in the disc. In both cases, with
young bars, one should see breaks in the radial abundance
distribution, which move outward in the disc as times evolves.
Some bulges, because they result in part from the stars in the bar being
scattered out of the galaxy plane, should be polluted by `young' stars
aged between 0.5 and 1.0 Gyr. Although one cannot exclude the 
presence of young stars in bulges, they are very difficult to detect.
The interstellar gas is a better laboratory to search for young bars.
The best candidate found so far for
a galaxy with a young bar is NGC 3359 where Martin \& Roy \shortcite{MR95}
have observed an abrupt O/H  gradient in the central region (which 
has an intense episode of star formation), and a flat gradient beyond
corotation. NGC 3319, observed by Zaritsky
et al. \shortcite{ZA94}, appears to be a case similar to NGC 3359.

In this paper we wish to explore the O/H radial distribution in
the prominent southern bar galaxy NGC 1365 in order to test 
the scenario of recent bar formation. NGC 1365 is as close to an archetype 
barred galaxy as one can find in the nearby universe. Its bar is very
rich in gas and it has moderate star forming activity.

\section{Observations and data reduction}

\subsection{The galaxy NGC 1365}

NGC 1365, a dominant galaxy of the Fornax cluster, is probably the most spectacular
of the nearby barred galaxies. Classified as SB(s)b I-II by de Vaucouleurs et al. 
\shortcite{dV91} (RC3), it has a Seyfert 1.5 type
nucleus and  a `hot spot' nuclear region
(Sersic \& Pastoriza \shortcite{SP65}; Anatharamaiah et al. \shortcite{AN93};
 Sandqvist et al. \shortcite{SA95}). A distance for NGC 1365 of
18 Mpc, as recently determined by a HST-WFPC2 Cepheid study  
\cite{MA96}, is adopted. The galaxy has clearly defined offset dust lanes 
on the leading edge of the bar and along its two main spiral arms. A  velocity jump has
been observed across the eastern and western dust lanes (\cite{LJ87}) indicative of strong shocks in the gas flow (Athanassoula \shortcite{AT92}).
The general properties of NGC 1365 are listed in Table 1.

\begin{table}
\caption{Global properties of NGC 1365} 
\begin{tabular}{lc}
\hline \hline
Parameter & Value \\ 
\hline 
$\alpha$ (J2000) & 3$^{\rm h}$33$^{\rm m}$37.$^{\rm s}$37 \\
$\delta$ (J2000) &  -36$^\circ$08$'$25.$''$5 \\
Morphological type & SB(s)b I-II\\
Inclination & 40$^\circ$\\
Position angle & 220$^\circ$ \\
Galactic extinction ($A_{\rm B})$ & 0.21\\
Systemic velocity (km s$^{-1}$) & 1640 \\
$\rho_{0}$  & 5.$'$61 \\
$b/a(i)$ & 0.51 \\
D (Mpc) & 18.2\\
Scale (pc arcsec$^{-1}$) & 88 \\
\hline
\end{tabular}\\
Notes: (a) Positions, angles and velocity
are from J\"ors\"ater \& van Moorsel \shortcite{JM95}; (b) type, extinction
and isophotal radius are from
de Vaucouleurs et al. \shortcite{dV91}; stellar bar axis
ratio corrected for inclination is from Martin \shortcite{MA95};
(d) Distance is from Madore et al. \shortcite{MA96}
\end{table}

Spectrophotometry of several \hii regions in the galaxy was obtained by
Pagel et al. \shortcite{PA79}, Alloin et al. \shortcite{AL81} and  
Roy \& Walsh \shortcite{RW88}; the
earlier observations, although limited, suggested relatively
high extinction, and a shallow global 
O/H abundance gradient, a now well-established feature of barred
galaxies (Vila-Costas \& Edmunds \shortcite{VE92}; Zaritsky et al. \shortcite{ZA94};
Martin \& Roy \shortcite{MR94}). Star formation activity is moderate or weak
in the bar, except in the nuclear ring (radius $\sim$ 7$''$).

NGC 1365 has been studied in \hi (e.g. Ondrechen \& van der Hulst \shortcite{OV89}), but
it is only recently that a detailed \hi investigation with the VLA  
has been completed and published by
J\"ors\"ater \& van Moorsel \shortcite{JM95}; these authors show that
NGC 1365 has a unique dropping rotation curve. Sandqvist et al. 
\shortcite{SA95} performed
radio continuum and CO observations of the central parts of the galaxy,
finding a radio jet and enhanced CO in the nucleus
and dust lanes; references to earlier radio work can also be found in
that paper. Sandqvist et al. \shortcite{SA95} estimated that the amount
of molecular gas in the nuclear and bar region, is equal to the
total amount of neutral atomic hydrogen in the galaxy,
that is 15 $\times$ 10$^9$ M$_\odot$.
The \hi distribution shows a hole in the central region where CO is the strongest and
the neutral hydrogen is predominantly located in the spiral arm.
The bar region is a very gas-rich feature.

\subsection{Selection of H~{\sc II} regions}

An AAT prime focus plate of NGC~1365 taken in the R band (RG 610 filter + 
O94-04 emulsion) on 1990 December 16 was scanned on the ESO PDS machine.
The triplet corrector was employed giving a scale of 15.$''$3 mm$^{-1}$ and the
scanning aperture was 50 $\mu$m and sampling 25 $\mu$m (0.$''$381). This
image is shown in Figure 1; 163 potential \hii regions were identified and
their X,Y positions determined by either fitting a 2-D Gaussian or more
often simply by centroiding the image by eye. Given that there was no
off-emission band image with which to check whether the identified regions
were indeed \hii regions, some contamination by stars, globular clusters
or distant galaxies was anticipated. In addition to the \hii regions candidates, 
the X,Y positions of HST GSC stars were measured by fitting 2-D Gaussians and a six
coefficient astrometric solution was made for 23 GSC stars. The RA and Dec
of the potential \hii regions was then calculated. 'Sky' positions were also
selected away from the galaxy and free of stars. In selecting \hii regions
to observe, a number of considerations were taken in account: 

\begin{enumerate}[(iv)]
\item about 55 object and sky fibres can be observed per fibre bundle per 
aperture plate; 
\item the full range of \hii region apparent brightness should be covered in 
order to avoid any bias; 
\item fibres cannot be closer than 19$''$. Thus only one \hii region could
be selected in a crowded region; a nearby \hii region could of course
be selected for a second aperture plate.
Some bright \hii regions should be included in common between
different aperture plates to check the observation quality and 
reproducibility of calibration; 
\item the faintest \hii regions could be included on more than one aperture plate
to increase the exposure time.
\end{enumerate}

\begin{figure*}
\centerline{\psfig{file=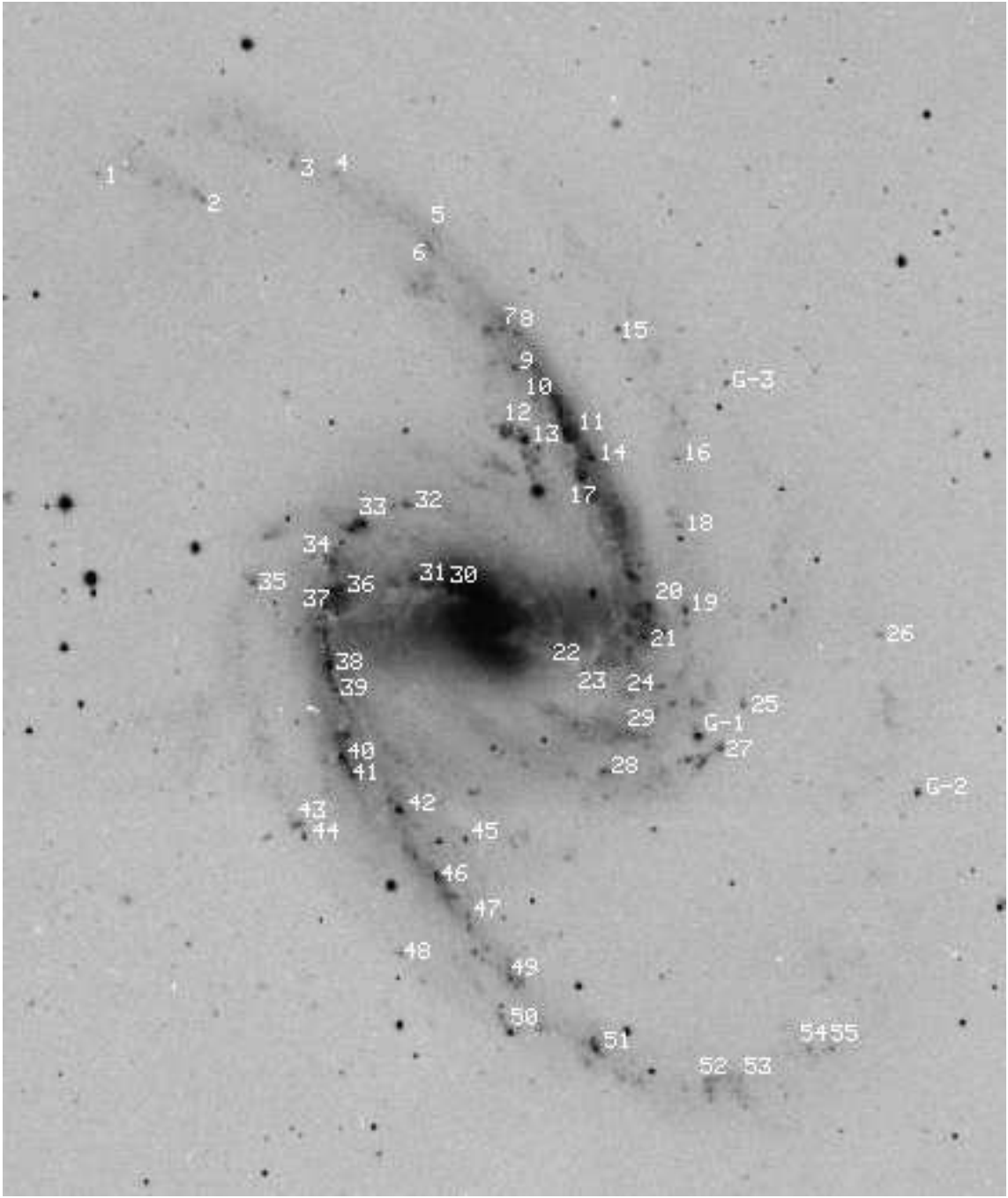,height=18.0cm,clip=}}
\caption{R band image of NGC 1365 obtained by David Malin
at the f/3.3 prime focus of the 3.9-m Anglo-Australian Telescope.
The numbers correspond to the \hii regions of Table 2; the
\hii regions are the brightest object seen closest to the number. G-1, G-2 and
G-3 are more distant galaxies discovered serendipitously. North is
up and east is left.}
\label{fig1}
\end{figure*}

  The final choice of \hii regions selected is indicated by the numbered
regions on Figure 1. These were divided between two aperture plates which
were observed on different nights. Both plate A (observed on 1994 December 09)
and plate B (observed on 1994 December 10 and 11) had 49 fibres on \hii regions 
and 7 on sky positions. There were 30 regions in common between the two plates:
8 had no detectable emission; 7 were so faint that the better exposure on
1993 December 11 only was used for analysis; the remaining 15 spectra
were combined: 5 had very faint emission and 5 were among the
brightest \hii regions observed. For targets not in common, 
9 objects were rejected which had insufficient signal-to-noise in the emission
lines. Three objects (G-1, G-2 and G-3 on Figure 1) were found to be
distant galaxies (See Appendix). Judging the 
strength of line emission from the appearance on a broad band image is clearly
not sufficient since some regions have stronger continuum, particularly
at smaller galactocentric radius.  The final number of distinct \hii region
spectra obtained with well measured lines was 55. 

\subsection{Fibre observations}
Low dispersion spectra of the selected targets in NGC 1365 were obtained
using the RGO spectrograph (25 cm camera) and the Tektronix \#2 1024 $\times$ 1024
CCD (24 $\mu$m pixels) on the 3.9 m Anglo-Australian Telescope during the nights of 1993
December 9-12. The FOCAP system with its bundle of about 55 fibres
of 300 $\mu$m core diameter (2.0$''$ on sky) was employed to 
feed the spectrograph; the FOCAP system was used because it allowed
closer positioning of fibres than AUTOFIB.
A 250 lines per mm grating was used in the
first-order blaze to collimator configuration for a dispersion of 156 \AA\ mm$^{-1}$.
Most of the \hii regions observed in NGC 1365 have
diameters in the range of 5$''$ -- 10$''$, much bigger
than the fibre apertures. To increase the area sampled by each fibre,
the telescope was set in a continuous scan motion, by moving it on a circle
of 1$''$ radius at a rate of one full circle per 30 s. Convolved
with seeing of 1$''$ -- 2$''$, the effective sampling area was
$\sim$ 3$''$ -- 4$''$ in diameter. The  Tektronix CCD is a thinned device and was used
in the XTRASLOW mode (394 sec readout time) to achieve the lowest readout noise 
of 2.3 e$^-$ rms.
The CCD  blue response is good, being  greater than 40\% at 3727 \AA. 

  Six 2000s exposures were taken with a first plate on 1993 December 09-10, one of which was
affected by cloud. A second aperture plate was used for observations on the
third night (1993 Dec 11-12) when conditions were good, seeing $\sim$1$''$,
and seven 2000s exposure were performed. The \hii regions indicated on Figure 1 are only those for which the emission
lines were analysed, the spectra containing at least detectable H$\alpha$ and 
H$\beta$ emission. Three of the objects rejected as having weak emission line spectra
were found to have high redshifts. They are indicated by a letter G prefix in
Figure 1 and are discussed in the Appendix.

\begin{figure*}
\centerline{\psfig{file=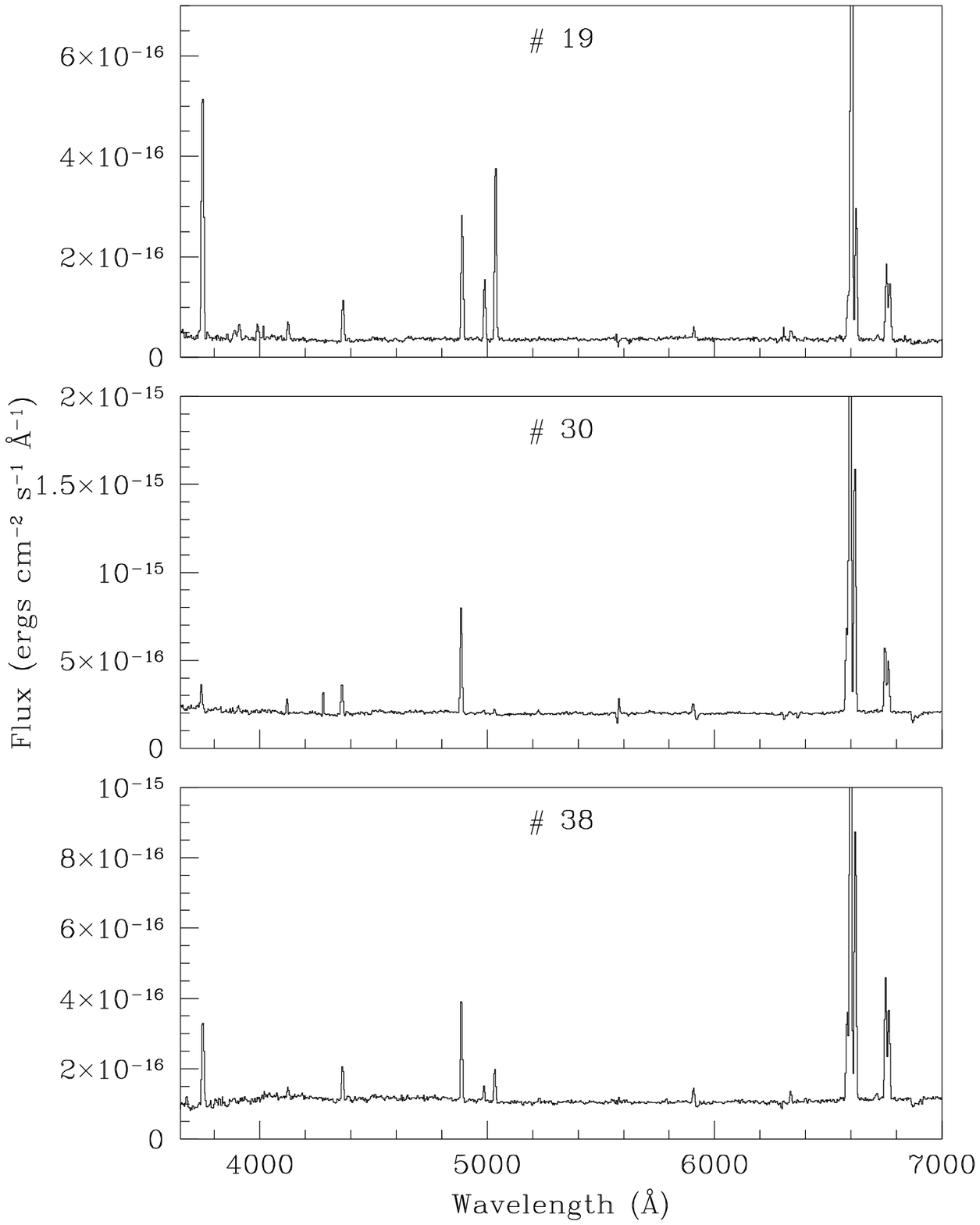,height=18.0cm,clip=}}
\caption{Examples of spectra of \hii regions in NGC 1365 obtained with FOCAP. 
The region numbers refer to those in Figure 1 and Table 2. No correction for interstellar 
extinction has been applied.}
\label{fig2}
\end{figure*}

\subsection{Spectrophotometry with optical fibers}

Spectroscopy using optical fibers for multiplexing is now a widely-used 
technique in astronomy. A great number of spectra are being
obtained but, unfortunately, little attention has been given
to the promising astrophysical potential of having usable spectral energy
distributions for so many sources in addition to radial velocities.
There is a general consensus that spectrophotometry cannot be done with fibres.
However in the multi-fibre spectroscopy of 49 \hii regions in NGC~2997 (Walsh \& Roy \shortcite{WR89}),
it was
shown that with suitable care, spectrophotometry to {\bf better} than 20\% can be 
achieved. Five problem areas for fibre spectrophotometry were outlined in 
Walsh \& Roy \shortcite{WR89}. The use of a CCD with high quantum efficiency,
as opposed to the IPCS for the NGC~2997 observations, together with the multi-night
coverage of one galaxy, have changed some of the emphasis of the problem areas
and led to improved understanding of the limitations. The problem areas can be
restated thus: dealing with differential atmospheric refraction; 
effect on spectra of cross talk between fibres; efficacy of sky subtraction; 
reliability of spectral extraction.

\begin{enumerate}[]

\item Atmospheric refraction

The fibres were 300$\mu$m in diameter (2.0$''$) so that at zenith distances
above 30$^\circ$ the differential atmospheric refraction between 3727 and 6730\AA\
(the extent of the most useful optical wavelength range for \hii regions) is 1.0$''$
(see Walsh \& Roy \shortcite{WR90}).
The maximum zenith distance at which exposures were made of NGC~1365 is 
40$^\circ$ so that differential refraction will have a modulating effect on
the spectrum. However the telescope was moved in a circular pattern of 1$''$
radius to improve the sampling of the extended \hii regions, and this also has the
effect of averaging out the effects of differential atmospheric refraction
for modest airmasses.

Comparison of spectra of the same \hii region taken at zenith distances of 6
and 35$^\circ$ and corrected for airmass, show only small differences in
line ratios between the blue and red ends. However for the observations
of the spectrophotometric standard stars (L745-46A and L870-2 \cite{OK74}) the
telescope was not circled during the exposure. Depending on the exact position of the
star image over the fibre and the airmass, then the input spectra will differ,
most obviously in the blue where the differential atmospheric refraction
changes rapidly with wavelength. As for the case for NGC~2997, four \hii
regions were in common between the fibre spectroscopy and the scanned long
slit spectroscopy (Roy \& Walsh \shortcite{RW87}). However for NGC~2997, the 
problems of relying on
the spectrophotometric standard for the flux calibration proved insurmountable
and a sensitivity curve was deduced by comparing the fibre spectra with the
data from a long slit over identical regions. Whilst the same method could have been
relied on for the NGC~1365 observations, it is clearly preferable to rely on an 
independent spectrophotometric calibration. 

  A number of precautions were taken in the observations of the standards.
At least three, and sometimes five, separate observations, each of 300s 
duration, were made on the standards and at low to moderate airmass (zenith
distance $\leq$ 27$^\circ$). At each exposure the standard star was recentered in the
guide fibre and then offset to the appropriate fibre. During data reduction 
the individual spectra of the standards were compared in
terms of absolute level and shape. The differences in absolute level
could be attributed to the star not being centered in the fibre combined with
the seeing, which was around 1$''$-1.5$''$. Differences
in spectral distribution were also apparent, in particular a relative in(de)crease
in the blue when comparing different exposures. This must be attributed to 
different centering of the star within the fibre: a higher blue response occurring
when the fibre was more centered to lower altitude zenith distance. 
Some conspiracy between shifts in the parallactic direction and 
perpendicular to it could produce such effects, although no attempt was made
to model this behaviour. 

In forming the mean spectrum of the standard for a
given night, a weighted mean of the strongest spectra was made; however if
a spectrum differed markedly in shape from the others, it was not included in the mean.
It was suggested in Walsh \& Roy \shortcite{WR89} that improved spectrophotometry
of standard stars could be achieved with very good seeing
by scanning the telescope in an elongated
pattern in the direction of the parallactic angle. This was not attempted during
the observations described here. Such a strategy will of course 
impair absolute spectrophotometry (which can only be achieved with good seeing
for fibres), but this is usually not a serious issue and
certainly was not of major concern here.
The agreement between spectra taken of the same \hii region on different nights
both with the same fibre and with a different fibre suggest that the strategy of 
scanning a small circle is sufficient to ensure repeatable spectrophotometry
to within 5\% and could also be applied to the observation of the standard
stars. Incidentally the level of agreement implies that the colour
transmission of the fibres is similar, at least to tolerances lower than the 
spectrophotometric ones.

  Inverse sensitivity curves were formed for L745-46A (first and third nights)
and for L870-2 (second, third and fourth nights). The two curves for L745-46A were
very significantly different in the blue, the curve derived for the third night
showing a sharp minimum centred at 4000\AA. For L870-2 the curves had the same general
shape but with differences in absolute level; for the second and third nights
the relative differences did not exceed about 6\%. The sensitivity curves
were used to calibrate the \hii region spectra and the extinction, derived from
comparison of the hydrogen line ratios with the Case B values, was computed.
It was found that the L870-2 sensitivity curves gave a more consistent 
value of the extinction (the Seaton \shortcite{SE79} Galactic curve
was employed as parametrized by Howarth \shortcite{HO83}) for the different 
hydrogen line ratios (e.g. H$\alpha$/H$\beta$ vs. H$\delta$/H$\beta$). It was 
therefore decided to use a mean weighted spectrum of
the L870-2 data from three nights to produce the calibration curve. On the
highest signal-to-noise \hii spectrum  (\#42, night 3) the 
extinction is the
same (within the errors of measurement) on all lines ratios from H$\alpha$/H$\beta$ to
H$\epsilon$/H$\beta$. Thus the calibration produced by the adopted L870-2 
spectrum is at least adequate over the range 3950 to 6600\AA. 

Comparison of the spectra with those of identical \hii regions
observed by Roy \& Walsh \shortcite{RW88} and Alloin et al. value \shortcite{AL81}
was also made. There are 9 \hii regions in common with Alloin et al
(L4, L6, L7, L10, L15, L21, L24, L28 and L33). There is poor agreement with
the IDS spectra of Alloin et al, where low extinctions ($c$=0) were derived; for the
IPCS spectra, for L33 the derived extinction is much higher than Alloin et al.
value, for L10 it is higher whilst for the others there is satisfactory 
agreement. The \oii/H$\beta$ ratios presented by Alloin et al. are 
systematically lower than presented here.
The spectra were also compared with those of Roy \& Walsh \shortcite{RW88} for 
\hii region emission summed over much larger areas than the fibre data.
The fibre data indicated higher extinction and higher dereddened [O~II]/H$\beta$ 
ratios than the summed \hii region emission. The higher extinction is most
probably a real effect whereby the bright cores have higher extinction, whilst
the higher \oii/H$\beta$ ratios may be affected by the
difficulties of fibre spectrophotometric calibration at lower wavelengths.

\item Cross-talk between fibres

  On account of the much higher signal-to-noise
of the CCD data in comparison with the previous IPCS data (Walsh \& Roy 
\shortcite{WR89}), a better
measurement of the cross-talk from fibre to fibre could be made. For an observation
of the standard star, the counts in the two adjacent fibres along the slit yielded
cross talk of 1.9 and 1.0\%, with an increase to 2.5\% in the red ($>$7000\AA) in
the fibre with the higher level of cross-talk. For the other fibre, with lower
cross talk, the spectrum had the same shape as that in the star fibre. A related
effect was noticed in that some spectra after flux calibration were very blue.
The effect appeared strongest on weaker spectra
although was not confined to them, and was also not simply correlated to the strength
of adjacent fibre spectra. It is suggested that the effect could arise
from the flat fielding, since the flat-field lamp has a low response in the blue.
This would make the flat field in the blue very sensitive to small changes
in packing between fibres along the slit and small irregularities would 
cause differential cross-talk. However it is not clear why this cross-talk should be
different between the flat field and the sky data. Sky flats might assist in
better understanding of this problem. It is clear that the fibres should be well
separated at the spectrograph entrance slit and on the detector in order to allow 
well-defined minima to be distinguished between the spectra.

\item Sky subtraction

As suggested by  Wyse \& Gilmore \shortcite{WG92}, at least six
fibres were employed for determination of the sky - in fact 7 were used. After 
normalising the
fibre to fibre response by the flat field spectra integrated in wavelength, the
mean sky spectrum was formed and subtracted from the object spectra. The efficacy of
sky subtraction was tested by examining the residuals on the mean sky-subtracted 
sky spectra. Examining the goodness of subtraction of the strong sky emission 
lines alone is not very reliable since small wavelength shifts ($\leq$0.3 pixels) 
between the fibre spectra can result 
in improperly subtracted sky lines, although the mean difference across the line
can be zero. Such small shifts can result from slight
differences in the polynomial fit to the comparison lines along the
slit. Comparison 
of the fibre-to-fibre response between the flat field and the [O~{\sc i}]5577\AA\
line shows good agreement, suggesting that the method of correcting for the
relative fibre transmissions is satisfactory.

\item Spectral extraction

 The packing of the fibres is such that the separation
of the spectra peak-to-peak is 5 or 6 CCD pixels, and there is not
a true zero intensity between fibres. The spectra were extracted simply by summing 
as many fibre spectra per object as possible; for the better separated spectra
one signal-free pixel column was left between fibre spectra. A full extraction
would require solving for all the extracted spectra simultaneously. Shifting the
extent of the extracted spectrum by one pixel causes an error of up to 5\%. 
Assigning a column from the edge of one spectrum into the adjacent spectrum
gives rise to an error of about 2\%. The simple linear extraction could be
improved by fitting the cross-dispersion profile and by optimal extraction, but
would not substantially affect the results presented here.

Comparing the spectra of the same \hii 
regions taken on different nights and with different fibres shows that these
systematic effects dominate the photon statistical errors associated with
the line fitting. For the bright lines, e.g. \oiii and H$\alpha$, the typical
differences in the line ratios (expressed as a fraction of H$\beta$) are 
10\%; for \oii this difference is around 20\%, with values up to 30\%.
For spectra taken on
different nights, but in the same fibre, the differences are only slightly larger
than the  photon statistical errors, although conditions were not photometric
on the first and second nights. It is clearly the problems of spectral
extraction and spectrophotometric calibration that dominate the uncertainties in
the derived line fluxes.

\end{enumerate}

\subsection{Reduction of the spectra}

All the data frames were bias subtracted and the spectra and flat fields formed
for each fibre. Wavelength calibration was achieved by fitting third order
polynomials to the integrated spectra of each fibre for a Th-Ar lamp. Corrections
of each individual exposure for airmass were made and the data summed. Following
flux calibration by the merged L870-2 spectrum, the spectrum of each target was
extracted and 
an interactive procedure for fitting continuum and emission lines was used
to derive the line fluxes $F(\lambda)$ expressed  in units of 
$F(H\beta)$ = 100. Photon noise errors on the fitted line fluxes were computed 
and propagated. The magnitude of the 
interstellar reddening was determined by the H$\alpha$/\Hb \ ratios;
comparison was done to  the
theoretical decrement as given by Brocklehurst \shortcite{BR71} for a temperature
of 8,000 K and a density of 100 cm$^{-3}$, but after adding 2 \AA\
of equivalent width to the \Hb \ emission line to compensate for
the underlying Balmer absorption (cf. McCall, Rybski \& Shields \shortcite{MC85};
Roy \& Walsh \shortcite{RW87}). If the underlying stellar absorption
at \Hb\ is greater than 2 \AA\, then this correction would lead
to an overestimation of the reddening correction, especially in \hii regions
of low \Hb\ emission equivalent width. The spectrum was corrected
in detail as a function of wavelength using the 
standard reddening law of Seaton \shortcite{SE79}, as
specified by Howarth \shortcite{HO83}, assuming R = 3.1.

\section{Results}

Figure 2 shows some examples typical of the best spectra, uncorrected for 
interstellar reddening. The region
numbers refer to the \hii region identification numbers in Figure 1
and Table 2. The difference in the relative strengths of the collisionally excited
lines of \oiii , \nii and \sii illustrate the change in excitation across
the galaxy. Region \#30 is close to the nucleus.
 Table 2 lists the line intensities (\Hb = 100) corrected for reddening for the
main emission lines of the 55 \hii regions observed. $X$ and $Y$ refer
to RA and DEC offsets in arcsec relative to the centre of the galaxy given in Table 1. The logarithmic
extinction at \Hb, $c$, includes both Galactic reddening (contribution about 0.07, 
see Table 1) and extragalactic extinction. $\rho/\rho_0$ is the radial distance
expressed in terms of the fractional isophotal radius (Table 1).
We made only a marginal detection of the WC9 star
in region \#30  (L4 of \cite{AL81}) found by Phillips \& Conti \shortcite{PC92};
this is in contrast to NGC~2997 where we observed 49 \hii regions
and found signatures of Wolf-Rayet stars in three (and possibly four) regions.

\subsection{Diagnostic diagrams}

Several line ratios have been calculated after correction for reddening.
Defining \oiii  as 1.34$\times I$\oiii 5007\AA, \nii
as 1.34$\times I$\nii 6584\AA\ and \sii as $I$\sii 6713$+$6730\AA, and using
\oiioiii as a sequencing parameter, Figure 3 shows \OiiOiii,
\niioii , \siioii and \oiiinii versus this parameter. The
tight relationships shown by these sequences imply that
the majority of the nebulae are ionization-bound; these trends
are characteristic of gas volumes photoionized by massive stars
(McCall et al. \shortcite{MC85}), and very similar, for example, to those measured in 
M101 (Kennicutt \& Garnett \shortcite{KG96}). The tight sequence shown by 
\niioii indicates that the
fibre spectrophotometry indeed produces reliable line fluxes.  The change
of \niioii and \siioii versus \oiioiii is driven by the thermal
properties of the \hii regions, and not by change in the nitrogen
and sulfur abundance
ratios (Garnett \& Shields \shortcite{GS87}). The abundance of oxygen mainly controls
the thermal equilibrium in \hii regions (McCall et al. \shortcite{MC85}).

\begin{table*}
\caption{H~{\sc ii} regions in NGC 1365 -- Reddening-corrected line fluxes (H$\beta$ = 100)}
\begin{tabular}{rrrrccccccc}
\hline \hline

RW &  X &    Y  &    [OII]  &    [OIII]    &    HeI  &      [NII] &   [SII]  &  c(H$\beta$)    & $\rho/\rho_0$ \\
\# & $''$ &  $''$ &   3727 &      5007    &     5876 &       6584   & 6717-30 &           &   \\
(1) & (2)   &  (3)  &    (4)    &   (5)     &   (6)       &   (7)    &  (8)   &     (9)   &     (10)  \\ 
\hline 
1 & 210  &  250  &  222$\pm$8 & 158$\pm$5 & 14.4$\pm$2.6 & 54$\pm$3 & 49$\pm$3 & 0.51$\pm$0.10 & 0.97 \\
2 & 151  &  235  &  368$\pm$17  &    70$\pm$4   &  12.9$\pm$3.0 &  81$\pm$4 &   80$\pm$4 & 0.66$\pm$0.13 & 0.84 \\
3 & 101  &   254  &   351$\pm$9  &     66$\pm$2 &  10.8$\pm$2.0 &  78$\pm$2  &  68$\pm$3 &  0.75$\pm$0.07 & 0.84 \\
4 &  76  &   248  &   606$\pm$51   &    63$\pm$9  &       & 108$\pm$10 & 137$\pm$12 & 0.80$\pm$0.25 & 0.81 \\
5 &  23  &   216  &   249$\pm$10  &    187$\pm$6  &    11.7$\pm$2.0 &  42$\pm$2  &  56$\pm$2 & 0.66$\pm$0.10 & 0.72 \\
6 &  26  &   207   &  364$\pm$26  &    100$\pm$8  &        &  99$\pm$7  &  90$\pm$9 & 0.74$\pm$0.18 & 0.68 \\
7 & -14   &  160   &  329$\pm$8  &   61$\pm$2  &    13.5$\pm$1.4 &  89$\pm$2  &  57$\pm$1 & 0.90$\pm$0.06 & 0.56 \\
8 & -22  &   158  &   476$\pm$28 &    45$\pm$4   &    &  98$\pm$6 &   71$\pm$6 & 0.97$\pm$0.16 & 0.56 \\
9 & -21  &   139   &  174$\pm$7  &     37$\pm$2  &    10.6$\pm$1.8 &  94$\pm$3 &   77$\pm$3 & 0.95$\pm$0.09 & 0.50 \\
10 & -31  &   131  &   262$\pm$14  &     35$\pm$4  &      & 103$\pm$5  &  79$\pm$5 & 0.89$\pm$0.12 & 0.48 \\
11 & -50  &   104  &   228$\pm$2   &    24$\pm$1   &    9.9$\pm$0.6 & 102$\pm$1  &  71$\pm$1 &  0.86$\pm$0.02 & 0.53 \\
12 & -14  &   106  &   222$\pm$5  &    123$\pm$3   &   10.7$\pm$1.1 &  73$\pm$2  &  49$\pm$1 &  0.84$\pm$0.06 & 0.37 \\
13 & -26  &    99  &   255$\pm$4  &     64$\pm$1  &    10.4$\pm$0.7 &  91$\pm$1  &  60$\pm$1 & 0.84$\pm$0.03 & 0.37 \\
14 & -63  &    88  &   221$\pm$6  &     82$\pm$2  &       & 71$\pm$2  &  51$\pm$2 & 0.73$\pm$0.06 & 0.42 \\
15 & -79  &   160  &   306$\pm$27  &  129$\pm$13  &     &  52$\pm$8  &  85$\pm$16 & 0.27$\pm$0.27 & 0.67 \\
16 & -111  &    87  &   264$\pm$25  &     53$\pm$9   &    &  80$\pm$8 &  114$\pm$12 & 0.58$\pm$0.25 & 0.55 \\
17 & -57  &    77  &   225$\pm$5  &     30$\pm$1   &    9.9$\pm$1.0 &  96$\pm$2  &  77$\pm$1 & 0.84$\pm$0.05 & 0.37 \\
18 & -112 &     50 &    614$\pm$18  &    115$\pm$4  &    12.6$\pm$2.3  & 84$\pm$3 &  105$\pm$5 & 0.77$\pm$0.08 & 0.47 \\  
19 & -114  &     2   &  366$\pm$5   &   134$\pm$2  &     9.6$\pm$0.7 &  65$\pm$1  &  63$\pm$2 & 0.76$\pm$0.04 & 0.41 \\
20 & -94  &     6   &  326$\pm$3   &    59$\pm$1  &    10.3$\pm$0.4 &  90$\pm$1  &  53$\pm$1 & 0.78$\pm$0.02 & 0.34 \\
21 & -91  &   -11  &   268$\pm$7  &     24$\pm$2  &    10.0$\pm$1.5 &  99$\pm$3  &  62$\pm$2 & 0.83$\pm$0.06 & 0.32 \\
22 & -41  &   -16  &   105$\pm$9  &     20$\pm$3  &     9.5$\pm$1.9 & 111$\pm$4  &  62$\pm$4 & 1.55$\pm$0.09 & 0.14 \\
23 & -57  &   -29  &   222$\pm$12  &     21$\pm$3  &     8.0$\pm$2.1 & 101$\pm$4  &  73$\pm$5 & 1.01$\pm$0.10 & 0.20 \\
24 & -89  &   -39  &   381$\pm$10  &     56$\pm$2  &     9.1$\pm$1.1 &  96$\pm$2  &  59$\pm$2 & 1.05$\pm$0.06 & 0.31 \\
25 & -146  &   -50  &   274$\pm$6  &    117$\pm$3  &    13.0$\pm$1.5 &  64$\pm$2 &   62$\pm$2 & 0.53$\pm$0.06 & 0.50 \\
26 & -223   &  -12   &  234$\pm$27  &    136$\pm$16   &      & 78$\pm$12 & 115$\pm$24 & 0.07$\pm$0.31 & 0.78 \\
27 & -133  &   -74   &  291$\pm$9   &   158$\pm$5   &   13.5$\pm$1.9 &  62$\pm$2  &  62$\pm$3 &  0.69$\pm$0.09 & 0.47 \\
28 & -67  &   -86  &   291$\pm$5   &    67$\pm$2    &  13.6$\pm$1.2 &  81$\pm$2  &  56$\pm$2 & 0.81$\pm$0.04 & 0.33 \\
29 & -81   &  -64  &   158$\pm$8  &     40$\pm$3   &      & 106$\pm$4 &   93$\pm$4 & 0.78$\pm$0.10 & 0.32 \\
30 &  16   &   15   &   48$\pm$2   &     4$\pm$.5   &      &  95$\pm$1&   38$\pm$2 & 1.19$\pm$0.02 & 0.07 \\
31 &  29  &    18  &   163$\pm$11  &     29$\pm$3   &       & 101$\pm$4  &  61$\pm$4 & 1.18$\pm$0.11 & 0.11 \\
32 &  40   &   64  &   354$\pm$13  &  34$\pm$3   &       & 107$\pm$4 &   91$\pm$3 & 0.87$\pm$0.08 & 0.23 \\
33 &  65   &   53 &    182$\pm$2  &     55$\pm$1  &    10.7$\pm$0.4 &  87$\pm$1 &   50$\pm$1 & 0.88$\pm$0.02 & 0.26 \\
34 &  93  &    35  &   269$\pm$6  &    119$\pm$3  &    16.0$\pm$1.5  & 71$\pm$2 &   51$\pm$2 & 0.47$\pm$0.06 & 0.32 \\
35 & 127   &   21 &    305$\pm$12  &     92$\pm$4  &     &  80$\pm$4 &   86$\pm$3 & 0.61$\pm$0.11 & 0.44 \\
36 &  78   &   17   &   80$\pm$3   &    14$\pm$2  &    8.7$\pm$1.2  & 96$\pm$2  &  57$\pm$2 & 0.67$\pm$0.06 & 0.27 \\
37 &  87   &    7   &  222$\pm$12  &     34$\pm$3   &       & 108$\pm$4  &  94$\pm$6 & 0.78$\pm$0.11 & 0.31 \\
38 &  85   &  -23  &   164$\pm$4   &    27$\pm$1  &     9.4$\pm$0.6 & 105$\pm$2 &   76$\pm$2 & 1.11$\pm$0.04 & 0.33 \\
39 &  81  &   -39 &    321$\pm$20   &    47$\pm$5   &   14.4$\pm$2.5 & 102$\pm$5 &   75$\pm$4 &  1.37$\pm$0.15 & 0.34 \\
40 & 78   &  -75   &  275$\pm$5 &      38$\pm$1  &     9.3$\pm$0.8 &  93$\pm$2 &   66$\pm$2 & 1.10$\pm$0.05 & 0.42 \\
41 &  72  &   -85  &   126$\pm$6   &    44$\pm$2   &   13.6$\pm$1.8 &  88$\pm$3  &  59$\pm$4 & 0.92$\pm$0.09 & 0.43 \\
42 &  47  &  -105  &   309$\pm$2   &    66$\pm$1  &    10.6$\pm$0.3 &  95$\pm$1  &  44$\pm$1 & 1.14$\pm$0.02 & 0.43 \\
43 & 104   & -113  &   516$\pm$10  &    122$\pm$3  &     9.1$\pm$1.6 &  74$\pm$3  & 104$\pm$3 & 0.53$\pm$0.06 & 0.60 \\
44 & 100  &  -120  &   276$\pm$8  &    240$\pm$6  &    13.3$\pm$2.3 &  52$\pm$3 &   59$\pm$4 & 0.40$\pm$0.08 & 0.61 \\
45 & 10  &  -122  &   227$\pm$5   &   103$\pm$2   &   11.1$\pm$1.7  & 76$\pm$2 &   59$\pm$2 & 0.61$\pm$0.05 & 0.42 \\
46 & 26  &  -144  &   357$\pm$9  &     82$\pm$3   &   13.3$\pm$1.7 &  82$\pm$2 &   61$\pm$4 & 0.63$\pm$0.07 & 0.52 \\
47 &  9  &  -165  &   374$\pm$20   &    91$\pm$6    &       & 89$\pm$5 &  109$\pm$7 & 0.58$\pm$0.14 & 0.57 \\
48 & 46  &  -185  &   324$\pm$13   &   107$\pm$5   &    9.9$\pm$2.8  & 73$\pm$4 &   80$\pm$5 & 0.63$\pm$0.11 & 0.69 \\
49 & -13 &   -200  &   541$\pm$53  &     63$\pm$9  &        &  78$\pm$9 &   81$\pm$12 & 0.80$\pm$0.29 & 0.67 \\
50 & -13 &   -230   &  428$\pm$4   &   228$\pm$2  &    11.5$\pm$0.4 &  44$\pm$1 &   35$\pm$1 & 0.69$\pm$0.02 & 0.77 \\
51 &  -61 &   -240  &   380$\pm$5  &     87$\pm$2   &    8.6$\pm$0.7  & 78$\pm$2  &  66$\pm$2 & 0.55$\pm$0.03 & 0.78 \\
52 & -123 &   -259  &   360$\pm$15  &    121$\pm$6   &   10.7$\pm$2.7 &  90$\pm$4  & 100$\pm$7 & 0.77$\pm$0.12 & 0.87 \\
53 &  -142  &  -258 &    180$\pm$12  &     76$\pm$6   &      & 86$\pm$6 &   76$\pm$7 & 0.34$\pm$0.16 & 0.89 \\
54 &  -181  &  -241  &   395$\pm$15  &    168$\pm$7   &      & 63$\pm$4  &  78$\pm$6 & 0.38$\pm$0.11 & 0.90 \\
55 &  -193  &   -242  &   532$\pm$51  &     60$\pm$9   &     &  77$\pm$9 &  110$\pm$13 & 0.74$\pm$0.28 & 0.92 \\
\hline
\end{tabular} 
\end{table*}

\subsection{Radial gradients in NGC 1365}

Galactocentric distances and azimuthal positions of the
\hii regions were calculated by assuming a 40$^\circ$ inclination of the galaxy disc to the 
plane of sky, and a position angle of 220$^\circ$ from J\"ors\"ater \& 
van Moorsel \shortcite{JM95}. As a normalizing radius, we used the value of
the isophotal radius $\rho_0$ of 5.$'$61 (RC3); the
possible choices of a normalizing radial parameter are discussed
in detail by Vila-Costas \& Edmunds \shortcite{VE92} and Zaritsky et al. \shortcite{ZA94}.
Line ratios which show systematic radial variations are
the excitation \ohb, the abundance indicators \niioiii and
\oiioiii and reddening (Fig. 4). The radial positions
of the \hii regions are given in Table 2 in terms of
the fractional isophotal radius. Despite a large scatter, there 
appears indeed to be a systematic
variation of $c$(\Hb), the logarithmic extinction at \Hb, as a function
of radius; such a radial trend has been observed so far only in the galaxy
M101 (Kennicutt \& Garnett \shortcite{KG96}). The linear coefficient
of correlation between $c$(\Hb) and $\rho/\rho_0$ is
R = --0.46. The value of $c$(\Hb)\ is high
on average across the disc, dropping from $\sim$1.2 near
the centre to 0.6--0.8 in the outer disc; the dust lanes of NGC 1365 are
remarkably strong and indicate the rich dust content. Despite the relatively high
extinction (the mean of $c$ = 0.75$\pm$0.27), one can see through the galaxy;
in particular, a distant elliptical  (object G-1 on Figure 1) was found just 
southwest of the bar.

\begin{figure*}
\centerline{\psfig{file=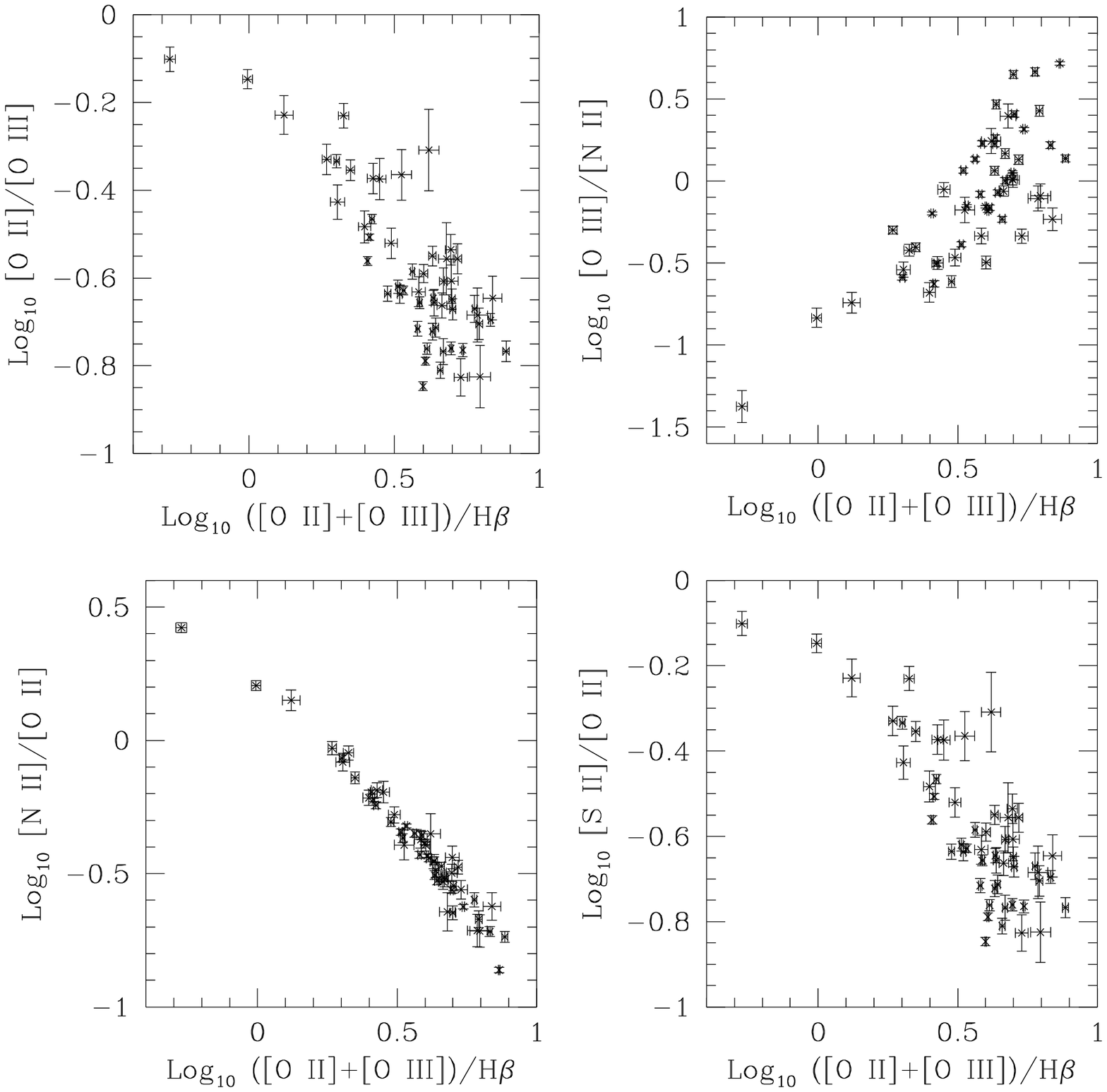,height=16.0cm,clip=}}
\caption{Diagnostic diagrams of log \OiiOiii, log \oiiinii, log \niioii
and  log \siioii vs. the sequencing index log \oiioiii are shown for the 
observed \hii regions in NGC 1365.}
\label{fig3}
\end{figure*}

As seen from the spectra of Figure 2, the level of excitation in NGC 1365
is generally low, and no direct measurement of the electron temperatures
is possible. Therefore semi-empirical methods must be relied on to derive abundances.
These methods have been widely discussed and used (e.g.
Edmunds \& Pagel \shortcite{EP84}; McCall et al. \shortcite{MC85}; Evans \shortcite{EV86};
Garnett \& Shields \shortcite{GS87}; Vilchez et al. \shortcite{VI88}; Walsh \& Roy 
\shortcite{WR89};
Belley \& Roy \shortcite{BR92}; Martin \& Roy \shortcite{MR94}; Zaritsky et al. 
\shortcite{ZA94}). They are based on the result that larger values of \oiioiii or 
\oiiinii are correlated with higher electron temperatures and lower abundances of 
oxygen. These calibrations have been refined by Edmunds and Pagel \shortcite{EP84},
McCall et al. \shortcite{MC85}, and Dopita \& Evans \shortcite{DE86}; Zaritsky et al.
\shortcite{ZA94} 
have produced a synthetic calibration for \oiioiii by merging the
three previous ones. It must be recalled that factors other than
heavy element abundances also modify nebular
line strengths. The limitations and uncertainties  of the
semiempirical methods due to various effects have been also discussed by
McGaugh \shortcite{MG91} for ionization parameter and effective temperature
of the ionizing radiation, by Henry \shortcite{HE93} and Shields \& Kennicutt
\shortcite{SK95} for dust, by Oey \& Kennicutt \shortcite{OK93} for density, by 
Kinkel and Rosa \shortcite {KR94}
for high Z effects and by Roy et al. \shortcite{RO96A} for moderately low Z behavior.

Radial gradients in \oiioiii and \oiiinii reflect primarily abundances
gradients wherever those line ratios are sampled over a significant
fraction of the \hii region volume. The absolute O/H abundance predicted by the different
semi-empirical calibrations of these ratio can vary depending on the 
calibration employed, but the relative trends can be probed reliably
(see for example Henry \& Howard \shortcite{HH95}; Zaritsky et al. \shortcite{ZA94}; 
Martin \& Roy \shortcite{MR94}; Kennicutt \& Garnett \shortcite{KG96}).
The radial abundance variation in oxygen abundances 
in NGC 1365 was determined
from three calibrations: i) \oiioiii as calibrated by Edmunds \& Pagel \shortcite{EP84};
ii) \oiiinii also calibrated by Edmunds \& Pagel \shortcite{EP84}; and
\oiioiii as calibrated by Zaritsky et al \shortcite{ZA94}. The reader can
then compare the results with similar abundance data on the few other
galaxies which have been thoroughly sampled with spectroscopic
techniques, e.g. NGC 2997 (Walsh \& Roy \shortcite{WR89}) and
M101 (Kennicutt \& Garnett \shortcite{KG96}). The overall trend in NGC 1365 is
a rather shallow abundance gradient at about --0.50 dex $\rho_0$$^{-1}$, 
or --0.02 dex/kpc;
the exact parameters of the correlation depends on the calibration used as shown
in Table 3, where $R$ is the coefficient of correlation.  The three calibrations, 
\oiiinii by Edmunds \& Pagel \shortcite{EP84}, \oiioiii by
Zaritsky et al. \shortcite{ZA94} and \oiioiii by Edmunds \& Pagel 
\shortcite{EP84}, give rather
similar results within the errors. The calibration of \oiioiii by Edmunds \& Pagel
\shortcite{EP84} tend to give a larger amplitude for the variation of O/H across 
the disc. A shallow gradient is expected for  strongly
barred galaxies, and NGC 1365 with a deprojected bar
axis ratio $b/a(i)$ = 0.51, has a strong bar 
(\cite{MA95}). The central intercept is at about 12 + log O/H = 9.15
(depending on the calibration used).

 In addition, one observes
a rather obvious break at R $\sim$ 185 - 200$''$ ($\rho = 0.55 - 0.60 \rho_0$); inside this
radius the O/H gradient is moderately steep (--0.80 dex $\rho_0$$^{-1}$, i.e. --0.05 dex kpc$^{-1}$), while
outside it is flat.  We have chosen $\rho = 0.55 \rho_0$ as the galactic
radial distance where the break occurs and have calculated the
appropriate equations and coefficient of correlation (Table 3);
this position for the  break ensures a reasonable number of sample points
(34 in the inner part and 21 in the outer part)
 while corresponding to small residuals for the fits. 
The correlations for the inner zone are almost as strong as
for all the points. Comparisons
between the three abundance calibrations can be made; it is obvious from Table 3, that
there is {\it no} abundance gradient (within the uncertainties) beyond the break.
We note that the values of O/H derived by using
the \oiiinii calibration by Edmunds \& Pagel \shortcite{EP84}
are very similar to those given by the synthetic calibration
of Zaritsky et al. \shortcite{ZA94}; this applies to individual
points and to the correlations (Table 3). The break is also well seen
in the radial plot of \ohb (not shown); thus
effects related to dust or a erroneous reddening determination  can therefore 
be excluded.

Possible azimuthal asymmetries in the O/H distribution were searched for,
but no evidence was found for such asymmetry. The significant assymetry in O/H
distribution found by
Kennicutt \& Garnett \shortcite{KG96} in M101 remains a rather unusual and
significant feature.
The apparent dispersion in abundance ($\sim$0.2 dex) at fixed radius is probably 
mainly due to variations in nebular ionization as discussed
by Kennicutt \& Garnett \shortcite{KG96}.
  
\begin{table*}
\caption{Equations of the O/H radial distribution in NGC 1365} 
\begin{tabular}{lcrc}
\hline \hline
Radius & Equation & $R$ & Calibration \\
\hline 
R $\leq$ 1.0 $\rho_0$ & 12 $+$ log O/H = 9.10$\pm$0.04 - 0.42$\pm$0.07 $\rho_0$ & -0.62 &\oiiinii (EP84)\\
R $\leq$ 1.0 $\rho_0$ & 12 $+$ log O/H = 9.17$\pm$0.05 - 0.53$\pm$0.09 $\rho_0$ & -0.63 & \oiioiii (ZA94)\\
R $\leq$ 1.0 $\rho_0$ & 12 $+$ log O/H = 9.12$\pm$0.06 - 0.68$\pm$0.11 $\rho_0$ & -0.65 &\oiioiii (EP84)\\
\hline \hline
R $\leq$ 0.55 $\rho_0$ & 12 $+$ log O/H = 9.23$\pm$0.06 - 0.78$\pm$0.17 $\rho_0$ &  -0.63 &\oiiinii (EP84)\\
R $\leq$ 0.55 $\rho_0$ & 12 $+$ log O/H = 9.26$\pm$0.08 - 0.77$\pm$0.21 $\rho_0$ & -0.54 & \oiioiii (ZA84)\\
R $\leq$ 0.55 $\rho_0$ & 12 $+$ log O/H = 9.28$\pm$0.10 - 1.10$\pm$0.28 $\rho_0$ & -0.57 & \oiioiii (EP84)\\
\hline
R $\geq$ 0.55 $\rho_0$ & 12 $+$ log O/H = 8.85$\pm$0.15 - 0.07$\pm$0.20 $\rho_0$ & -0.09 & \oiiinii (EP84)\\
R $\geq$ 0.55 $\rho_0$ & 12 $+$ log O/H = 8.66$\pm$0.17 + 0.13$\pm$0.23 $\rho_0$ & 0.13 & \oiioiii (ZA84)\\
R $\geq$ 0.55 $\rho_0$ & 12 $+$ log O/H = 8.50$\pm$0.18 + 0.16$\pm$0.24 $\rho_0$ & 0.15 & \oiioiii (EP84)\\
\hline 
\end{tabular}\\
\end{table*}

\begin{figure*}
\centerline{\psfig{file=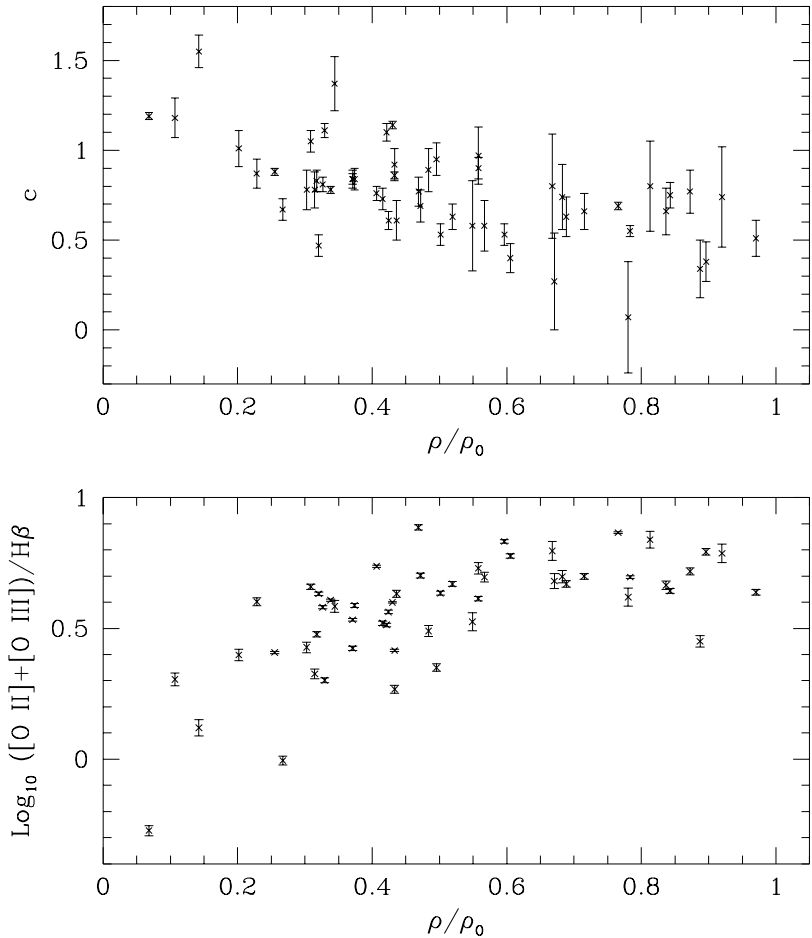,height=14.0cm,clip=}}
\caption{(a) The extinction $c$(\Hb) is plotted against the normalized isophotal
radius in NGC 1365. (b) The sequencing index log \oiioiii vs. the normalized
isophotal radius is shown.}
\label{fig4}
\end{figure*}

\section{Discussion}

\begin{figure*}
\centerline{\psfig{file=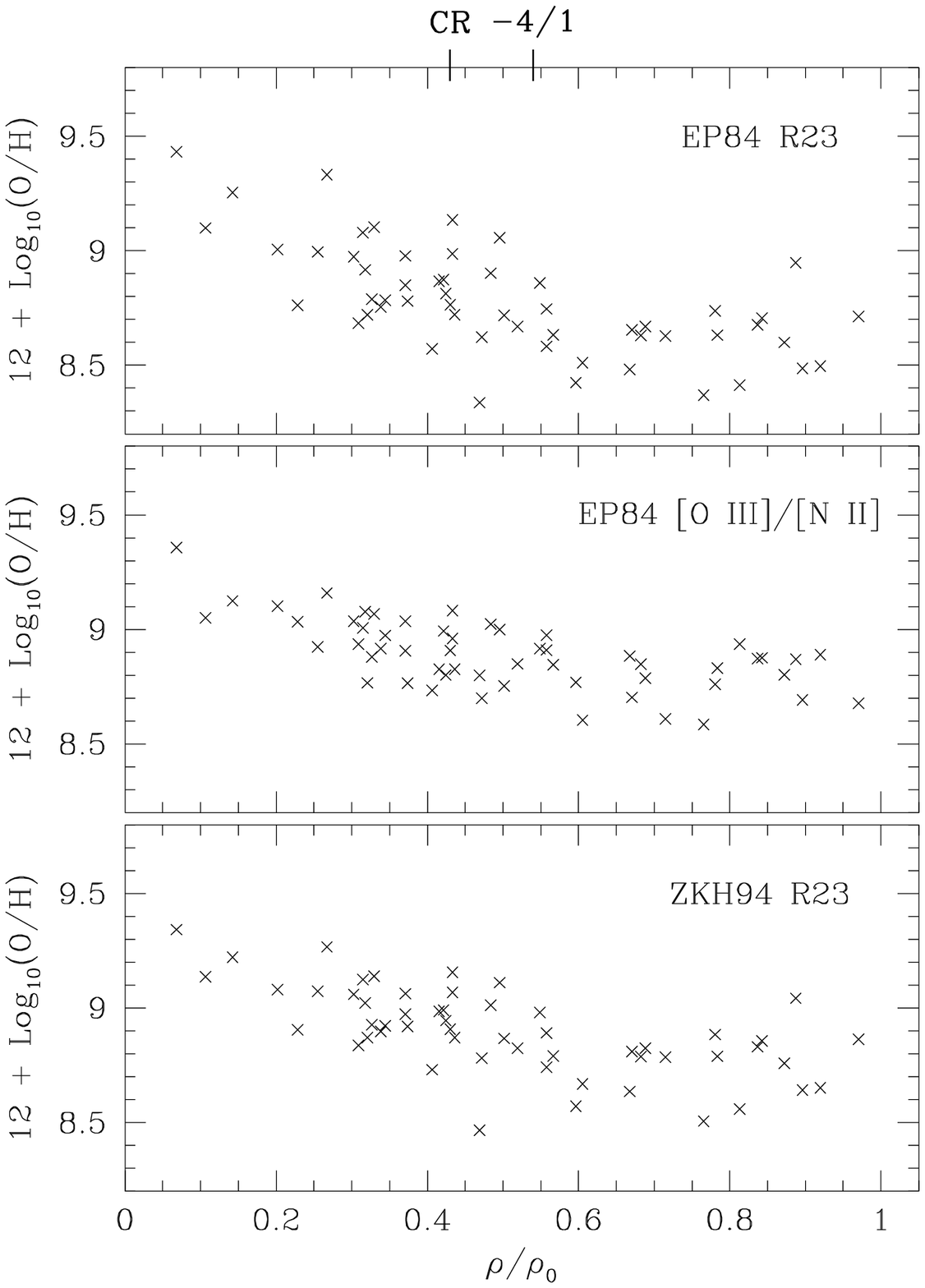,height=16.0cm,clip=}}
\caption{(a) The gradient in oxygen abundance across NGC 1365 using the
the calibration of \oiioiii by Edmunds \& Pagel (1984). (b) The
same gradient using the calibration of \oiiinii by Edmunds \& Pagel
(1984). (c) The gradient in oxygen abundance using the synthetic
calibration of \oiioiii by Zaritsky et al. (1994). CR and -4/1 indicates the
radial positions of the galaxy corotation  and -4/1 resonance respectively
as found by J\"ors\"ater \& van Moorsel (1995).}
\label{fig5}
\end{figure*}

\begin{figure*}
\centerline{\psfig{file=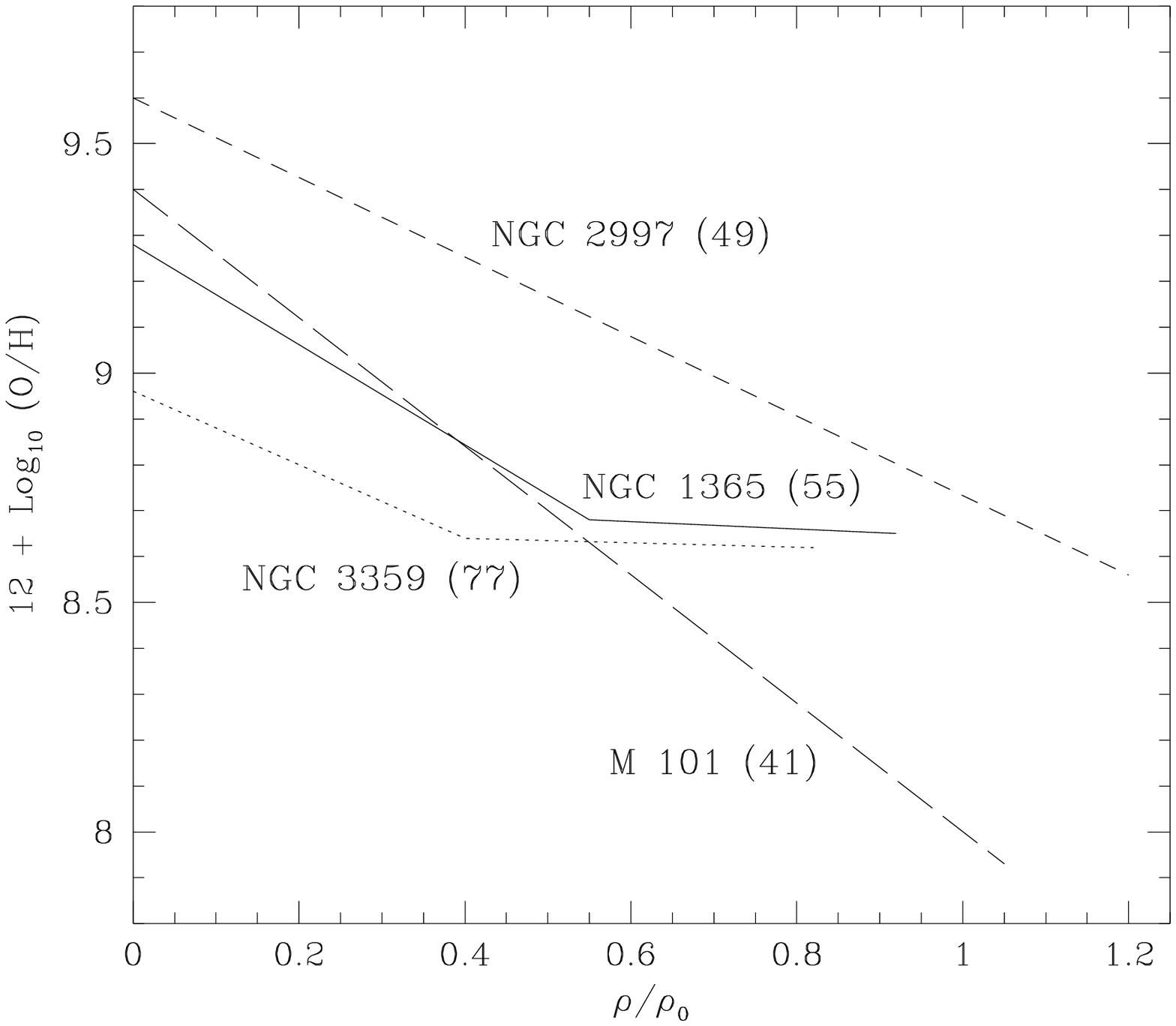,height=10.0cm,clip=}}
\caption{ Comparison of the O/H abundance gradients in the interstellar gas
of two barred galaxies (NGC 1365 and NGC 3359) and two
normal spirals (M 101 and NGC 2997). The
number in parenthesis indicates the sample size of  \hii regions
in each galaxy.}
\label{fig6}
\end{figure*}

The presence of a break in the radial abundance
distribution is one outstanding signature that a bar
may have recently ($\leq$ 1 Gyr) formed in a spiral galaxy. 
Following Friedli et al. \shortcite{FR94} and
Friedli \& Benz \shortcite{FB95} who have
developed this scenario, two breaks may actually appear
in the radial abundance distribution: (i) a ``steep-shallow''
break, near corotation, due to vigorous enrichment by
star formation in the bar, followed by a much flatter portion
dominated by the dilution effect of the outward
flow, and (ii) an outer ``shallow-steep'' break defining
where the outflow has had time to penetrate 
the outer disc. This ``shallow-steep'' outer
break moves to larger radius as time evolves,
and as the extent of the radial homogenization reaches further out
in the disc (Friedli et al. \shortcite{FR94}; Friedli \& Benz \shortcite{FB95}).
As the bar ages, it dissolves and runs out of gas. Star formation then
weakens, and dilution effects in the interstellar gas come to dominate;
after a few Gyr,
the resultant global abundance gradient ends up shallow, the break
becomes indistinguishable and the gradient displays
a monotonic decrease. The slope of any
pre-existing radial gaseous or stellar abundance gradient  decreases
from the start of the first phase. To have a break, one needs vigorous
star formation in the bar, and grand design spiral arms for
strong outward flows.
NGC 1365 is a barred galaxy with moderate star formation along the bar
but an intense nuclear starburst (Seyfert nucleus and
ring of \hii regions); this is evidence for an age of $\sim$1 Gyr
(see discussion in Martin \& Roy \shortcite{MR95}).

 A ``steep-shallow'' inner break is found in the abundance
distribution of NGC 1365 and it is located quite a bit further out
than the corotation radius (Figure 5).
It is assumed that due to an interaction or a merger,
the disc of NGC 1365 became unstable and developed a bar in a
recent epoch.
On account of angular momentum transfer via the arms, the bar induced large-scale
radial flows of interstellar gas. First,
a quick phase of intense star formation occured along the
bar major axis as well as along spiral arms; galaxies
showing this property (e.g. NGC 3359) have rather young bars. 
At a later time (closer to our epoch), a starburst was triggered when
gas had collapsed to the centre in a nuclear
ring as observed by Sandqvist et al. \shortcite{SA95}. Star formation now continues 
mainly along the spiral arms.
With the fuel  progressively consumed, star formation
is essentially limited to the ring and the spiral arms.  
Thus in this framework of a recently formed bar,
 NGC 1365 would be a disc spiral with a strong,
gas rich, but moderately young bar (age $\sim$1 Gyr). As indicated by the absence of a rich
population of \hii regions in the bar, the star formation
rate is now weak, except in the center and at the bar ends. 

 From the neutral hydrogen
kinematics, corotation (at $r = 145''$ or at $\rho \sim 0.43 \rho_0$)
is at 1.21 times the bar semi-major axis
(see Linblad, Linblad \& Athanassoula \shortcite{LLA96}.
There is a clear break in the O/H abundance radial distribution 
at $\rho \sim 0.55\rho_0$ or $r = 185''$ (Figure 5). Although
there is some uncertainties ($\pm 10\%$) in the locations of corotation
and of resonances (dependence on the model)
and of the break, the latter is beyond corotation; 
the break in the O/H gradient is close to the -4/1
resonance at $r = 180''$. The position of the abundance gradient break
$\sim 30''$ beyond corotation is puzzling, since this
is a region of strong radial mixing. However mixing always competes
with enrichment due to star formation in changing the local abundance.
The intense star forming activity in NGC 1365 observed at the bar ends and 
just beyond (betrayed by clumps of several bright \hii regions) may
compensate dilution and maintain high abundances, pushing the
location of the break further where star formation is reduced.

The apparent dispersion of the O/H points, at
a fixed radius, does not vary with galactocentric distance.
This is to be contrasted with the behavior of another
barred galaxy, NGC 3359, which has also a break in 
its radial abundance distribution (see Fig. 6). NGC 3359 has vigorous star
formation in its bar and  a steep inner abundance gradient; 
there the break is observed at
corotation. The abundance fluctuations, at a fixed radius, are much
larger in the outer parts of the disc -- where the gradient is flat --
than in the inner regions where the gradient is steep.
Martin \& Roy \shortcite{MR95} have interpreted this behavior 
of the O/H abundances in NGC 3559
as due to a recently formed  bar $\sim$400 Myr old, an age less than
the mean azimuthal mixing time in galaxy discs \cite{RK95}. 
We suggest that the homogenizing action of the bar in NGC 1365 may have acted on a timescale longer
than the characteristic azimuthal mixing time, that is the
bar is older than 500 Myr. The high mass of gas accumulated in the centre,
as implied by the presence of a nuclear starburst and a Seyfert nucleus,
also indicates an age $\geq$ 1 Gyr for the bar of NGC 1365. Overall,
several features are consistent with a bar formed
about 1 Gyr ago.

In Figure 6, the abundance gradients of four galaxies which
have had extensive spectrophotometry across their discs are shown:
two normal spiral galaxies, NGC 2997 (Walsh \& Roy \shortcite{WR89}) and M 101
(Kennicutt \& Garnett \shortcite{KG96}); two barred 
galaxies, NGC 3359 (Martin \& Roy \shortcite{MR95}) and NGC 1365 (this work).
The shallower gradients and the presence of breaks in the radial
abundance distribution is
easily seen in the two barred galaxies. The difference between
barred and normal spirals in radial
abundance distributions is striking. If the recently formed bar hypothesis
is correct, the numerical simulation of Friedli et al. \shortcite{FR94}
of a disc with a recently formed bar and star formation
show that the pre-bar O/H radial distribution in NGC 1365 would have 
been very similar to that observed nowadays in M~101.

Finally, the models of Friedli and collaborators
predict two breaks in the radial abundance distribution of galaxies
with young bars. So far no observational evidence
for the ``shallow-steep'' break in the outer disc has been found.
In order to ascertain the scenario of secular evolution of
disc galaxies (Martinet \shortcite{MAR95}), by the recurrent action of bars, it will be
necessary to find this outer break. This more definite test of the ``young bar'' hypothesis appears to be difficult however. As 
gas density falls with radial distance and the star formation
rate weakens, there are fewer and smaller \hii regions,
not unlike the interarm \hii regions. Long integration times will be required
in order to determine O/H abundances and  there may be only a few galaxies having a sufficient number of outer
\hii regions to provide statistically significant samples.

\noindent
{\bf Acknowledgments} \\
We acknowledge the technical support of J. Pogson, the AAT telescope 
operator. Discussions with Daniel Friedli, Pierre Martin, and
Laurent Drissen were most helpful.
We thank the PATT Committee for assigning time on the AAT for this
project. This investigation was funded in part by the Natural
Sciences and Engineering Research Council of Canada, the Fonds FCAR
of the Government of Quebec and by
the Visitor Program of the European Southern Observatory through financial
support of JRR.

\vspace{1cm}

\noindent
{\Large \bf Appendix A: Serendipitously discovered galaxies} \\

As mentioned in section 2.2 three of the targets chosen as \hii regions were 
identified from their spectra as galaxies. They are indicated as G-1 to G-3 on
Figure 1. Brief details of the targets and their spectra are included here.

\noindent
{\bf G-1:} This bright galaxy is quite round (ellipticity on the B image $\sim$ 0) so 
is probably of type around E0-E1. Its major axis has a half width of about 4$''$,
corrected for seeing. The spectrum shows a strong red continuum with absorption
lines of Ca~{\sc II} H and K, H$\gamma$ and G-band and Mg~{\sc I}. From the wavelengths
of the Ca~{\sc II} H and K lines the red-shift is 0.103. There is a 
detection of Ca~{\sc II} H and K absorption at the redshift of NGC 1365, making the
galaxy a very useful probe for studying the line of sight velocity dispersion
in an interesting region near the bar.

\noindent
{\bf G-2:} This galaxy has a resolved almost circular core of half width $\sim$2$''$
and shows two spiral arms of total extent $\sim$15$''$. The spectrum shows
a redward rising continuum with Ca~{\sc II} H and K and Mg~{\sc I} absorption lines.
From the observed wavelength of the Ca~{\sc II} H and K lines, the redshift is 0.294.

\noindent
{\bf G-3:} This galaxy displays a strong disc with a possible knot at its SE edge,
possibly indicating an interacting system. The spectrum shows a rather flat continuum
with strong emission lines of \oii, H$\beta$, and \oiii. The redshift is
0.131. The \oiii 5007/H$\beta$ ratio is 2.4 and \oii/\Hb\ ratio is 4.5.


\begin{thebibliography}{}


\bibitem[\protect\citename{Alloin et al. }1981]{AL81} Alloin D.,
 Edmunds M. G., Lindblad P. O., Pagel B. E. J., 1981, A\&A, 101,
377

\bibitem[\protect\citename{Anantharamaiah et al. }1993]{AN93}
Anantharamaiah K. R., Zhao J.-H., Goss W. M., Viallefond F., 1993
ApJ, 419, 585

\bibitem[\protect\citename{Athanassoula }1992]{AT92} 
Athanassoula E., 1992, MNRAS, 259, 345


\bibitem[\protect\citename{Belley \& Roy }1992]{BR92} Belley J., Roy J.-R., 1992,
 ApJS, 87, 61

\bibitem[\protect\citename{Brocklhurst}1971]{BR71}
Brocklehurst, M., 1971, MNRAS, 153, 471

\bibitem[\protect\citename{Combes \& Elmegreen}1993]{CE93}
 Combes F., Elmegreen B. G., 1993, A\&A, 271, 391

\bibitem[\protect\citename{de Vaucouleurs et al. }1991]{dV91} 
de Vaucouleurs G., de Vaucouleurs A., Corwin H. G., Buta, R., Paturel G.,
Fouqu\'e R., 1991, Third Reference Catalogue of Bright Galaxies.Springer, New York
(RC3)

\bibitem[\protect\citename{Dopita \& Evans }1986]{DE86}
Dopita M. A., \& Evans I. N., 1986, ApJ, 307, 431 

\bibitem[\protect\citename{Edmunds \& Pagel }1984]{EP84}
 Edmunds M. G., Pagel B. E. J., 1984, MNRAS, 211, 507

\bibitem[\protect\citename{Evans }1986]{EV86}
Evans I. N., 1986, ApJ, 309, 544

\bibitem[\protect\citename{Friedli et al. }1994]{FR94} 
Friedli D., Benz W., Kennicutt R. C., 1994, ApJ, 430, L105 

\bibitem[\protect\citename{Friedli \& Benz }1995]{FB95} 
Friedli D., Benz W., 1995, A\&A, 301, 649


\bibitem[\protect\citename{Garnett \& Shields}1987]{GS87} 
Garnett D., Shields G. A., 1987, ApJ, 317, 82


\bibitem[\protect\citename{Henry }1993]{HE93}
Henry R. B. C., 1993, MNRAS, 261, 306

\bibitem[\protect\citename{Henry \& Howard }1995]{HH95}
 Henry R. B. C., Howard J. W., 1995, ApJ, 438, 170

\bibitem[\protect\citename{Howarth }1983]{HO83}
Howarth I. D., 1983, MNRAS, 203, 301


\bibitem[\protect\citename{J\"ors\"ater \& van Moorsel }1995]{JM95} 
J\"ors\"ater J., van Moorsel G. A., 1995, AJ, 110, 2037

\bibitem[\protect\citename{Kennicutt \& Garnett }1996]{KG96} 
Kennicutt R. C., Garnett D. R., 1996, ApJ, 456, 504


\bibitem[\protect\citename{Kinkel \& Rosa }1994]{KR94}
Kinkel U., Rosa M. R., 1994, A\&A, 282, 37

\bibitem[\protect\citename{Lindblad \& J\"ors\"ater }1987]{LJ87}
Lindblad P. O., \& J\"ors\"ater S., 1987, in Evolution of Galaxies,
ed. J. Palous. Tenth European Regional Astronomical Meeting of the
IAU, p. 289

\bibitem[\protect\citename{Lindblad et al. }1996a]{LI96}
Lindblad P. A. B., Lindblad P. O., Athanassoula E., 1996,
in Barred Galaxies, IAU Coll. no. 157, ed. R. Buta, B. Elmegreen,
D. Crocker. ASP Conference Series, p. 413

\bibitem[\protect\citename{Linblad et al. }1996b]{LLA96}
Lindblad P. A. B., Lindblad P. O., Athanassoula E., 1996,
A\&A, 313, 65

\bibitem[\protect\citename{Madore et al. }1996]{MA96}
Madore B., et al., 1996, BAAS, 189, 108.04


\bibitem[\protect\citename{Martin }1995]{MA95}
Martin P., 1995, AJ, 109, 2428

\bibitem[\protect\citename{Martin \& Roy }1994]{MR94} 
Martin P., Roy J.-R., 1994, ApJ, 424, 599

\bibitem[\protect\citename{Martin \& Roy }1995]{MR95}
 Martin P., Roy J.-R., 1995, ApJ, 445, 161

\bibitem[\protect\citename{Martinet }1995]{MAR95}
 Martinet L., 1995, Fund. Cosm. Phys., 15, 141

\bibitem[\protect\citename{McCall et al. }1985]{MC85}
 McCall M. L., Rybski P. M., Shields G. A., 1985, ApJS, 57, 1

\bibitem[\protect\citename{McGaugh }1991]{MG91} McGaugh S. S., 1991, ApJ, 380, 140

\bibitem[\protect\citename{Noguchi }1988]{NO88} Noguchi M., 1988, A\&A, 203, 259

\bibitem[\protect\citename{Noguchi }1996a]{NO96A}
Noguchi M., 1996,  in Barred Galaxies, IAU Coll. no. 157, ed. R. Buta, B. Elmegreen,
D. Crocker. ASP Conference Series, p. 339

\bibitem[\protect\citename{Noguchi }1996b]{NO96B} Noguchi M., 1996, ApJ, October in press

\bibitem[\protect\citename{Norman \& Sellwood }1996]{NS96}
Norman, C. A., Sellwood J. A., 1996, ApJ, 462, 114

\bibitem[\protect\citename{Oey \& Kennicutt}1993]{OK93} 
Oey M. S., Kennicutt R. C., 1993, ApJ, 411, 137

\bibitem[\protect\citename{Oke }1974]{OK74}
Oke J. B., 1974, ApJS, 27, 21


\bibitem[\protect\citename{Ondrechen \& van der Hulst }1989]{OV89} 
Ondrechen M. P., van der Hulst, J. M., 1989, ApJ, 342, 29

\bibitem[\protect\citename{Pagel et al. }1979]{PA79}
 Pagel B. E. J., Edmunds M. G., Blackwell D. E., Chun M. S.,
Smith G., 1979, MNRAS, 189, 95

\bibitem[\protect\citename{Pfenniger }1992]{PF92} Pfenniger D., 1992, in Physics
 of Nearby Galaxies, Nature or
Nurture?, ed. T. X. Thuan, C. Balkowski, J. Tran Tranh Van. Ed. Fronti\`eres,
Gif-sur-Yvette, p. 519

\bibitem[\protect\citename{Phillips \& Conti }1992]{PC92}
Phillips A. C., Conti P. S., 1992, ApJ, 395, L91

\bibitem[\protect\citename{Roberts et al. }1979]{RO79}
 Roberts W. W., Huntley J. M., van Albada G. D., 1979, ApJ, 233, 67

\bibitem[\protect\citename{Roy \& Walsh }1987]{RW87}
Roy J.-R., Walsh J. R., 1987, MNRAS, 228, 883

\bibitem[\protect\citename{Roy \& Walsh }1988]{RW88}
 Roy J.-R., Walsh J. R., 1988, MNRAS, 234, 977

\bibitem[\protect\citename{Kunth \& Roy }1995]{RK95}
Roy J.-R., Kunth D., 1995, A\&A, 294, 432

\bibitem[\protect\citename{Roy et al. }1996a]{RO96A}
 Roy J.-R., Belley J., Dutil Y., Martin P., 1996, ApJ, 460, 284

\bibitem[\protect\citename{Roy}1996b]{RO96B}
Roy J.-R., in Barred Galaxies, IAU Coll. no. 157, ed. R. Buta, B. Elmegreen,
D. Crocker. ASP Conference Series, p. 63

\bibitem[\protect\citename{Sandqvist et al. }1995]{SA95}
 Sandqvist Aa., J\"ors\"ater S., Lindblad P. O., 1995, A\&A, 295, 598

\bibitem[\protect\citename{Scowen et al. }1992]{SC92}
 Scowen P. A., Dufour R. J., Hester J. J., 1992, AJ, 104, 92

\bibitem[\protect\citename{Seaton }1979]{SE79}
Seaton M. J., 1979, MNRAS, 187, 73P

\bibitem[\protect\citename{Sellwood \& Wilkinson }1993]{SW93}
Sellwood J. A., Wilkinson A., 1993, Rep. Prog. Phys., 56, 173

\bibitem[\protect\citename{Sersic \& Pastoriza }1965]{SP65}
Sersic J. L., Pastoriza M., 1965, PASP, 77, 287

\bibitem[\protect\citename{Shields \& Kennicutt }1995]{SK95}
Shields J. C., Kennicutt R. C., 1995, ApJ, 454, 807

\bibitem[\protect\citename{Shlosman \& Noguchi}1993]{SN93}
Shlosman I., Noguchi M., 1993, ApJ, 414, 474

\bibitem[\protect\citename{Tubbs}1982]{TU82} Tubbs A. D., 1982, ApJ, 255, 458

\bibitem[\protect\citename{Vila-Costas \& Edmunds }1992]{VE92} 
Vila-Costas M. B., Edmunds M. G., 1992, MNRAS, 259, 121

\bibitem[\protect\citename{Vilchez et al. }1988]{VI88}
Vilchez J. M., Pagel B. E. J., D\'{\i}az A. I., Terlevich E.,
Edmunds M. G., 1988, MNRAS, 235, 633

\bibitem[\protect\citename{Walsh \& Roy }1989]{WR89} 
 Walsh J., Roy J.-R., 1989, ApJ, 341, 722

\bibitem[\protect\citename {Walsh \& Roy }1990]{WR90}
Walsh J. R., Roy J.-R., 1990, in 2nd ESO/ST-ECF Data Analysis
Workshop. ed. D. Baade and P. J. Grosbol. ESO, Garching, p. 95

\bibitem[\protect\citename{Walsh \& Roy }1996]{WR96}
Walsh J. R., Roy J.-R., 1996, MNRAS, submitted

\bibitem[\protect\citename{Wyse \& Gilmore }1992]{WG92}
Wyse R. F. G., Gilmore G., 1992, MNRAS, 257, 1

\bibitem[\protect\citename{Zaristsky et al. }1994]{ZA94}
 Zaritsky D., Kennicutt R. C., Huchra J. P., 1994, ApJ, 420, 87

\end{thebibliography}
\end{document}